\documentclass[12pt,a4paper]{article}
\usepackage{autobf,axodraw,epsfig,reportno}
\epsfclipon
\def\Strut  {\rule[0pt]{0pt}{2ex}}
\def\eg     {\emph{e.g.}}
\def\etc    {\emph{etc.}}
\def\cf     {\emph{cf.}}
\def\ie     {\emph{i.e.}}
\def\etal   {\emph{et al}} % the period must be added by hand
\def\MSbar  {\overline{\rm MS}}
\reportno{EPTCO-98-004}
\title{Theory of Spin Effects in\\
       Hard Hadronic Reactions\thanks{Invited review talk, presented
         at The {XIII} International Symposium on High Energy Spin
         Physics, Protvino, Russia, September 8-12, 1998.}}
\author{Philip G. Ratcliffe\\
  \em Dip. di Fisica, Universit\`a dell'Insubria\thanks{The
    \emph{Insubri} were a Celtic population originally from across the
    Alps, who settled the Canton Ticino and the northern part of the
    region now known as Lombardy in the V century B.C., founding the
    city now called Milan early in the IV century B.C.},\\ 
  \em via Lucini 3, 22100 Como, Italy\\ 
  and\\ 
  \em Ist. Naz. di Fisica Nucleare---Sezione di Milano\\ 
  E-mail: pgr@fis.unico.it}
\date{Oct.~1998}
\begin{document}
\maketitle
%---------------------------------- Abstract ----------------------------------%
\begin{abstract}
  I discuss the present situation with regard to a variety of
  theoretical topics in hadronic spin physics: (a) global analysis of
  the $g_1$ data---positivity at leading and next-to-leading order,
  renormalisation-scheme dependence, parametrisation, and hyperon
  $\beta$-decays; (b) items from the realm of transverse
  spin---twist-three effects, single-spin asymmetries, and
  transversity; and finally (c) recent developments in understanding
  the $Q^2$ evolution of orbital angular momentum.
\end{abstract}
\newpage
%--------------------------------- Main Body ----------------------------------%
\section{Introduction}

Let me begin by thanking the Local Organising Committee for kindly
inviting me to attend the Symposium and for large amount the time and
effort that went into making my arrival in Protvino at all possible. I
should also like to take this opportunity to congratulate them for
succeeding, notwithstanding the obvious difficulties, in attaining a
balanced mix of informative and instructive talks from both
experimentallists and theorists alike.

%------------------------------------------------------------------------------%
\subsection{General Outline}

Given the generously broad title assigned to me, I have attempted to
at least touch upon those subjects not already covered by other
plenary or parallel speakers. From the schedule, it emerged that the
following areas were among those least represented in the theoretical
talks (although some have been partially covered in the experimental
presentations):
\begin{itemize} \itemsep0pt \parskip0pt
\item \underbar{global $g_1$ data analysis}---renormalisation scheme
  dependence, positivity at LO and NLO, and hyperon $\beta$-decays;
\item \underbar{transverse spin}---twist-three, single-spin
  asymmetries, transversity, and inequalities;
\item \underbar{orbital angular momentum}---evolution and gauge
  dependence.
\end{itemize}
Thus, following a few introductory notes on factorisation formul\ae\ 
and global spin sum rules, I shall endeavour to give a flavour of the
present stage of development of the above topics. That said, the use
and importance of inequalities in transverse-spin variables will only
fully emerge as and when precise data become available. Moreover, the
problem of gauge invariance in orbital angular momentum is
particularly technical and perhaps of little phenomenological
significance as yet. Therefore, while recently arousing increasing
theoretical interest, these last two topics will not be covered here.
Finally, to avoid unnecessary repetition, I shall omit detailed
definitions wherever possible, which may be found in either earlier
talks or the original literature.

%------------------------------------------------------------------------------%
\subsection{Factorisation Formul\ae}

Schematically, the cross-section for the hadronic process $AB\to{}CX$ is
\begin{equation}
\label{eq:factform}
  F_A(x_A)    \; \otimes \;
  F_B(x_B)    \; \otimes \;
  d\hat\sigma \; \otimes \;
  D_C(z_C),
\end{equation}
where $F_i(x_i)$ are partonic densities, $d\hat\sigma$ is the partonic
hard-scattering cross-section, and $D_i(z_i)$ is a fragmentation
function. The symbols $\otimes$ represent convolutions in $x_i$ and
$z_i$, the partonic longitudinal momentum fractions; and there is an
implicit sum over parton flavours and types. Each term in the above
expression has an expansion in both $\alpha_s$ and
twist.\footnote{Twist may usefully be viewed as simply a convenient
  labelling or ordering of the power-suppressed contributions in the
  asymptotic limit.}

Cross-section (\ref{eq:factform}) simplifies considerably in certain
cases: \eg, when one or more of the partons is replaced by a photon
(or weak boson), or if the final state is unobserved and is therefore
to be summed over. It is also important to recall that spin does not
represent an obstacle to the factorisation procedure nor to
application of the above formula: the quantities relating to polarised
particles are merely replaced by their spin-weighted counterparts
(single-spin asymmetries are slightly more involved, requiring some
form of angular weighting).  It is instructive to recall the following
aspects of the formula:
\begin{itemize} \itemsep0pt \parskip0pt
\item radiative corrections induce logarithmic scale dependence in all
  factors (expressed via an $\alpha_s$ expansion);
\item factorisation is carried out ``twist-by-twist'';
\item it is already more complicated at twist 3, in that diagrams
  \emph{na{\"\i}vely} higher-order in $\alpha_s$ can contribute even
  at leading order;
\item twist-3 cross-sections are constructed with one and only one of
  the terms calculated at twist 3; the rest are calculated at twist 2,
  as usual.
\end{itemize}

The third point above is a common source of error: na{\"\i}vely, one
might expect twist-3 effects to be due only to explicitly
higher-dimension terms, \eg, the quark mass. However, it is now known
that the dominant twist-3 contributions come from diagrams with an
extra partonic leg,\cite{Twist3} associated with an \emph{apparent}
extra power of $\alpha_s$.  Moreover, relations involving twist-2
contributions require that the factor of $\alpha_s$ be absorbed into
the correlation functions,\cite{Twist3} thus promoting such
contributions to truly leading order in $\alpha_s$.  Hence, the
\emph{only} suppression asymptotically is the typical $M/p_T$
associated with twist 3, which means that one might reasonably
\emph{expect} such effects to be large: \eg, even for $p_T\sim10$\,GeV
(assuming the natural mass scale $M$ to be of the order of the nucleon
mass) the asymmetries should be of order 10\%.

A complication now emerging, with the realisation that large twist-3
single-spin asymmetries (SSA) may exist, is that there are many
possible sources. At next-to-leading twist (\ie, three), all terms but
one in the above factorisation product are taken at leading twist
(two) and just one term at the next contributing twist (three).  Thus,
we are faced with the problem of isolating the true source among
several possibilities, which might all turn out to contribute.

%------------------------------------------------------------------------------%
\subsection{Global Sum Rules}

Another important and intuitive decomposition is that of the $z$-axis
projection of the total nucleon spin:
\begin{equation}
  J_z^p
  = \frac12
  = \frac12 \Delta \Sigma + \Delta g + L_z^{q+g},
\end{equation}
together with the twin sum rule for the transverse projection:
\begin{equation}
  J_T^p
  = \frac12
  = \frac12 \Delta_T \Sigma + \Delta_T g + L_T^{q+g}.
\end{equation}
I include the transverse-spin sum rule merely as a reminder of its
existence. There are extra subtleties here: for example, the
densities, $\Delta_T\Sigma$, have twist-3 contributions (absent for
longitudinal polarisation).

Difficulties in the definitions of partonic densities are caused by
both scheme and gauge dependence:
\begin{description} \itemsep0pt \parskip0pt
\item{(\emph{i})\hphantom{i}} renormalisation ambiguities mix
  $\Delta\Sigma$ and $\Delta{g}$ at NLO;
\item{(\emph{ii})} the separation into spin and orbital components is
  gauge dependent.
\end{description}
To some extent, the problem of gauge dependence is circumvented by the
natural axial-gauge choice in factorisation proofs and formul\ae.
However, the problem of identifying operators with meaningful physical
quantities is fraught with ambiguity. Much attention has recently been
paid to the orbital angular momentum case; \cite{GI-OAM} for lack of
space the reader is referred to the literature.

%------------------------------------------------------------------------------%
\section{Global \protect\autobf{$g_1$} data analysis}
%------------------------------------------------------------------------------%
\subsection{Positivity in Parton Densities}

The experimental asymmetry is expressed (at leading twist) as
\begin{equation}
  A_1
  \equiv
  \frac{\sigma_{1/2}-\sigma_{3/2}}
       {\sigma_{1/2}+\sigma_{3/2}}
  =
  \frac{g_1(x,Q^2)}{F_1(x,Q^2)}.
\end{equation}
Thus, $g_1$ is bounded by $F_1$: $|g_1|{\leq}F_1$. Now, at the
partonic level, $F_1$ and $g_1$ are defined in terms of sums and
differences of helicity densities:
\begin{equation}
  f = f^\uparrow+f^\downarrow,
  \qquad
  \Delta f = f^\uparrow-f^\downarrow.
\end{equation}
Therefore, the positivity of $f^{\uparrow,\downarrow}$ \emph{would}
lead to a useful bound:
\begin{equation}
  |\Delta f(x,Q^2)| \leq f(x,Q^2).
\end{equation}
However, beyond LO there is no guarantee of positivity; even quark
helicity flip is possible (via the ABJ axial anomaly).

In this respect, note that, \emph{in principle}, even the
\emph{un}-polarised densities could become negative (owing to virtual
corrections). Only in the na{\"\i}ve parton model (or LO) are the
densities positive definite (by definition). At NLO, the ambiguities
inherent in the choice of renormalisation scheme make negative
densities possible in particular schemes. To understand this, recall
that \emph{physical} quantities correspond to partonic densities
multiplied by coefficient functions (a power series in $\alpha_s$);
beyond LO, partonic densities themselves should not be thought of as
physically measurable quantities. In fact, positivity is partially
rescued by the fact that if higher-order corrections became large
enough to change the sign, perturbation theory would not be valid.

%------------------------------------------------------------------------------%
\subsection{Positivity Beyond LO}

\begin{figure}[htb]
\epsfig{figure=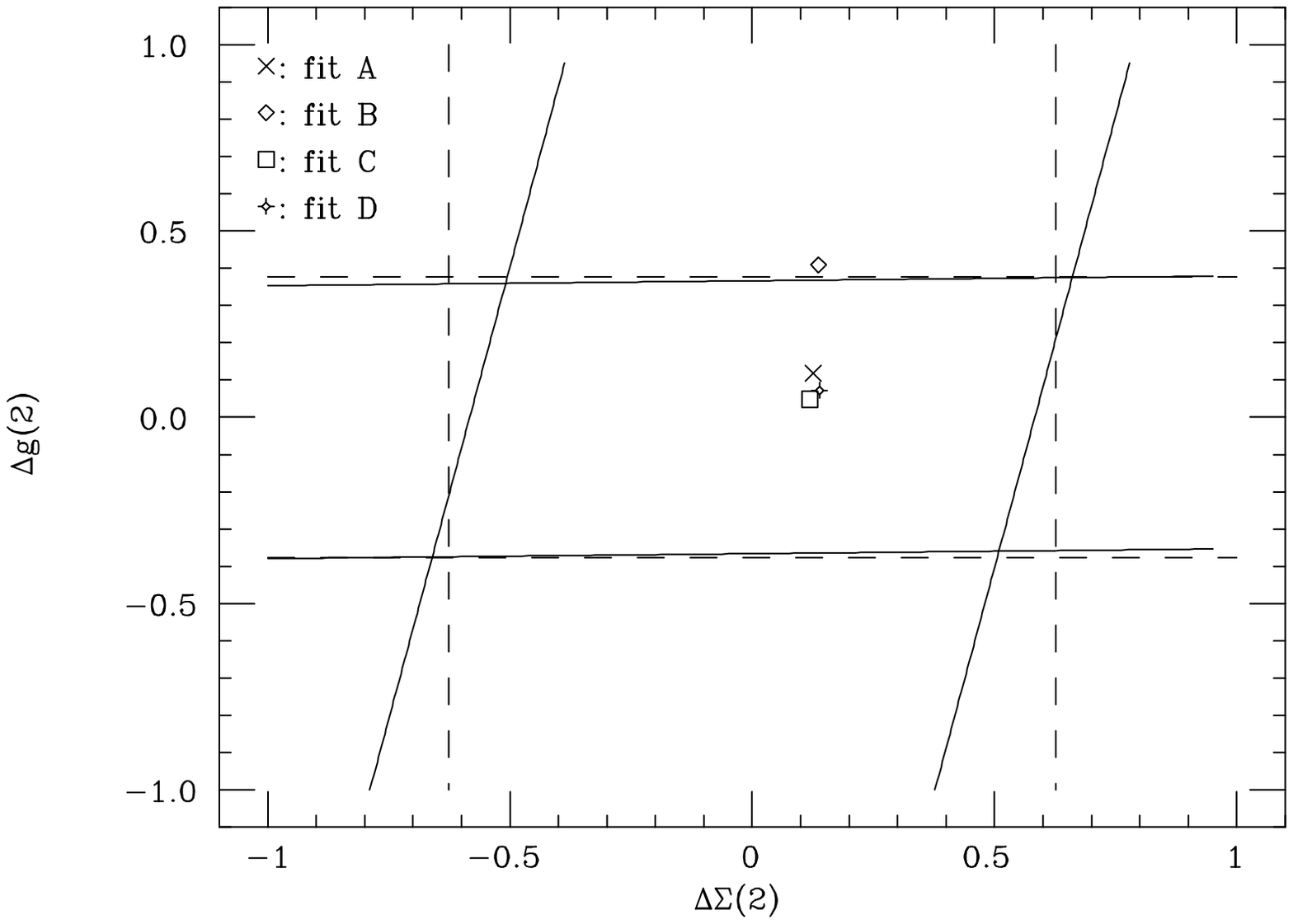,width=55mm}\hfill
\epsfig{figure=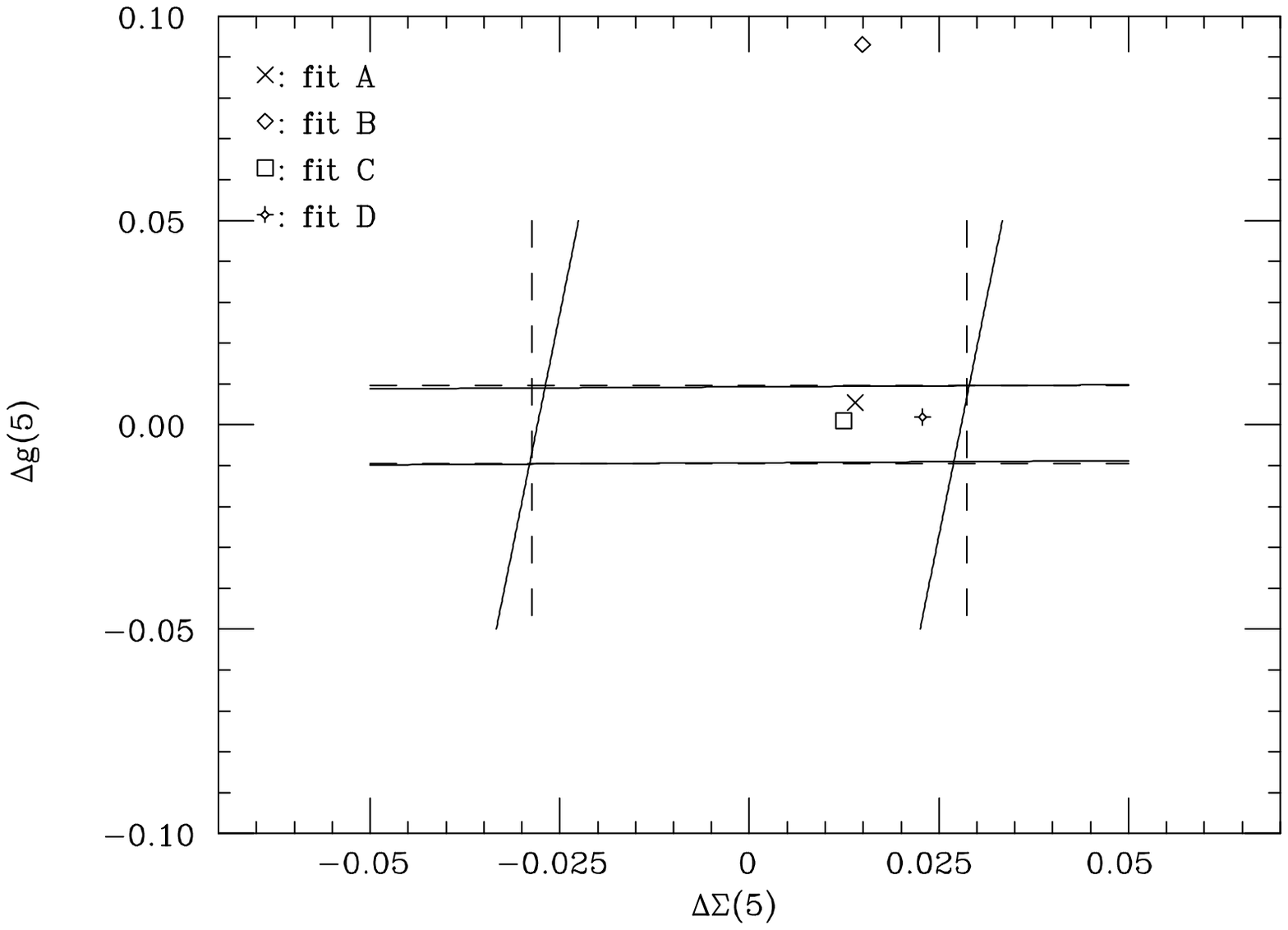,width=55mm}\relax
\caption{\label{fig-far2}
  The LO (dashed lines) and NLO (solid lines) positivity bounds on
  $\Delta\Sigma(N)$ and $\Delta{g(N)}$ for $Q^2=1$\,GeV$^2$ and $N=2$,
  5, from Altarelli \etal.\protect\cite{Altarelli:1998nb}}
\end{figure}

\begin{figure}[htb]
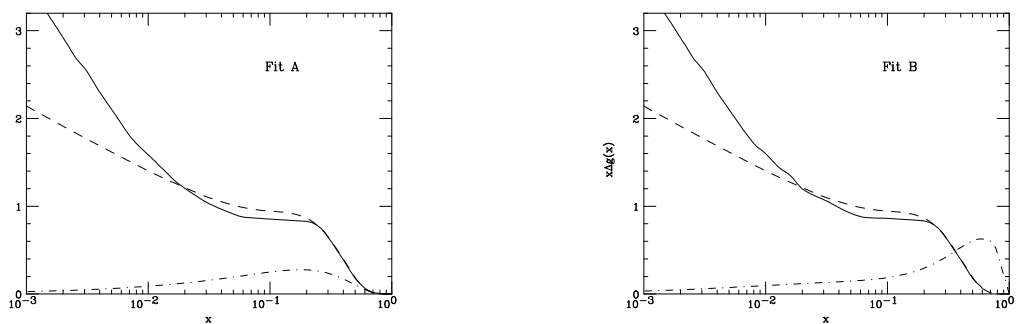

\epsfig{figure=fig-far3a.eps,width=55mm}\hfill
\epsfig{figure=fig-far3b.eps,width=55mm}
\caption{\label{fig-far3}
  The maximal gluon density at $Q^2=1$\,GeV$^2$ obtained from LO
  (dashed lines) and NLO (solid lines) positivity bounds, using
  polarized quark densities from two fits of Altarelli
  \etal.\protect\cite{Altarelli:1996nm} The corresponding best-fit
  polarized gluon density is also shown (dot-dashed).}
\end{figure}

Including the NLO corrections, the inequalities take on the following
form (moment-by-moment but suppressing $N$ and $Q^2$ for
clarity): \cite{Altarelli:1998nb,Forte:1998x1}
\begin{equation}
  \frac{
  \left| 
  \left( 1 + \frac{\alpha_s}{2\pi} \Delta C^d_\Sigma \right) \Delta\Sigma
           + \frac{\alpha_s}{2\pi} \Delta C^d_g              \Delta g
  \right|
  }
  {
  \left( 1 + \frac{\alpha_s}{2\pi}        C^d_\Sigma \right) \Sigma
           + \frac{\alpha_s}{2\pi}        C^d_g              g
  }
  \leq 1,
\end{equation}
using DIS as the natural \emph{defining} process for quark densities.
And
\begin{equation}
  \frac{
  \left| 
  \left( 1 + \frac{\alpha_s}{2\pi} \Delta C^h_g \right) \Delta g
           + \frac{\alpha_s}{2\pi} \Delta C^h_\Sigma    \Delta \Sigma
  \right|
  }
  {
  \left( 1 + \frac{\alpha_s}{2\pi} C^h_g \right) g
           + \frac{\alpha_s}{2\pi} C^h_\Sigma    \Sigma
  }
  \leq 1,
\end{equation}
using Higgs production as a possible \emph{defining} process for the
gluon density.\cite{Altarelli:1998nb,Forte:1998x1} This is actually a
\emph{gedanken} experiment, in which one imagines producing a Higgs
particle via a gluon-proton collision. The bounds so derived are shown
for two example moments in fig.~\ref{fig-far2}.  Such bounds may be
useful to pin down the shape of $\Delta{g(x)}$, see
fig.~\ref{fig-far3}.
At present, $\Delta{g(x)}$ is essentially determined via scaling
violations alone, which fix only the low moments with any precision,
since $|\gamma_{qg}|\ll|\gamma_{qq}|$ for large $N$ (see
fig.~\ref{fig-far4}):
\begin{figure}[htb]
\centering
\epsfig{figure=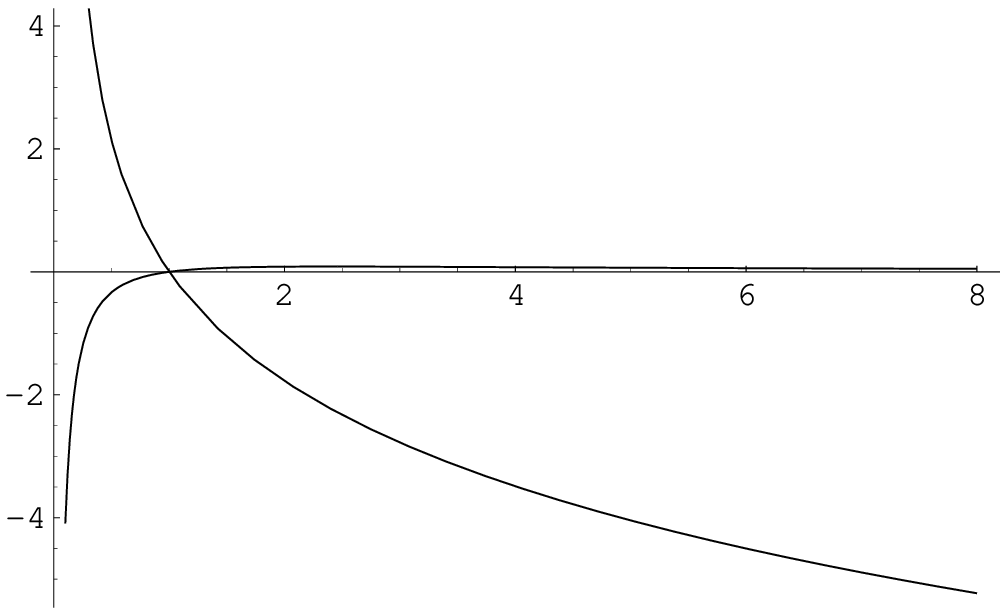,width=55mm}
\caption{\label{fig-far4}
  The LO anomalous dimensions $\gamma_{qq}(N)$ and $\gamma_{qg}(N)$
  (respectively, the top and bottom curves at small $N$) as a function
  of $N$, from Forte \etal.\protect\cite{Forte:1998x1}}
\end{figure}
\begin{equation}
  \frac{d}{dt} \, g_1^{\rm singlet}
  =
  \frac{\langle e^2\rangle}{2} \frac{\alpha_s}{2\pi}
  \left[\Strut
   \gamma_{qq} \Delta\Sigma + 2n_f \gamma_{qg}\Delta g
  \right]
  + \mathrm{O}(\alpha_s^2).
\end{equation}

%------------------------------------------------------------------------------%
\subsection{More on Positivity}

Analysis of the evolution of individual spin components shows the
problem to be partially ``self-curing''; \cite{Bourrely:1998x1} at LO
the IR-singular terms (with the usual $+$ prescription) lead to
\begin{equation}
  \frac{dq(x)}{dt} =
  \frac{\alpha_s}{2\pi}
  \left[
   \int_x^1 dy \frac{q(y)}{y} P\left(\frac{x}{y}\right)
   - q(x) \int_0^1 dz P(z)
  \right].
\end{equation}
The second term cannot change the sign of $q(x)$ as it is diagonal in
$x$: as $q(x)$ approaches zero, so too does the very term driving it
toward the sign change. Full $q$-$g$ mixing leads to (for example)
\begin{eqnarray}
  \frac{dq_+(x)}{dt}
  &=&
  \frac{\alpha_s}{2\pi}
  \left[
    P_{++}^{qq}\left(\frac{x}{y}\right) \otimes q_+(y)
  + P_{+-}^{qq}\left(\frac{x}{y}\right) \otimes q_-(y)
  \right.
\nonumber\\
  & &
  \quad
  \left.
  + P_{++}^{qg}\left(\frac{x}{y}\right) \otimes g_+(y)
  + P_{+-}^{qg}\left(\frac{x}{y}\right) \otimes g_-(y)
  \right].
\end{eqnarray}
Again, the only singular terms in this and the three companion
equations are diagonal (in parton type too) and therefore cannot spoil
positivity.

The result survives to NLO order,\cite{Bourrely:1998x1} except for a
small violation in the $qg$ and $gq$ kernels at $x\sim0.7$, which, it
is conjectured, might be cured by an appropriate choice of $\gamma_5$
scheme. Thus, positivity may be a useful addition to the data fitters'
armoury. However, care is required to avoid those schemes in which it
could cause undesirable bias in fit results.

%------------------------------------------------------------------------------%
\subsection{Renormalisation Scheme Choice}

In order to analyse data, a certain amount of theoretical input is
necessary.  Thus, there are several other issues (some of which are
apparently exquisitely theoretical) requiring careful examination
since they can in fact have a significant impact on the outcome of
global data fits involving parton evolution:
\begin{itemize} \itemsep0pt \parskip0pt 
\item definition of polarised gluon and singlet-quark densities,
\item small-$x$ extrapolation,
\item choice of initial parametrisation.
\end{itemize}

Though mathematically acceptable, a peculiarity of the $\MSbar$ scheme
is that some soft contributions are included in the Wilson coefficient
functions, rather than being absorbed into the parton densities.
Consequently, the first moment of $\Delta\Sigma$ is not conserved and
it is difficult to compare the DIS results on $\Delta\Sigma$ with
constituent quark models at low $Q^2$.  To avoid such oddities, Ball
\etal.\cite{Ball:1996td} have introduced the so-called Adler-Bardeen
(AB) scheme, now a common choice.  The AB scheme involves a minimal
modification of the $\MSbar$ scheme; the polarised singlet quark
density is fixed to be scale independent at one loop:
\begin{equation}
  a_0(Q^2) \; = \;
  \Delta\Sigma \, -  \, n_f \frac{\alpha_s(Q^2)}{2\pi} \, \Delta g(Q^2).
\end{equation}
Other factorisation schemes will not alter $\Delta{g(Q^2)}$ greatly
but may, in contrast, cause $\Delta\Sigma$ to vary considerably. As a
result, the values of $a_0(\infty)$ and $\Delta\Sigma$ will be very
different.  Recall that $\Delta{g(Q^2)}$ grows as $1/\alpha_s(Q^2)$.
Of course, the difference between any two schemes lies in the
(unknown) higher-order terms.  Thus, comparison of results between two
schemes (\eg, AB and $\MSbar$) could also shed light on the importance
of the NNLO corrections.

Zijlstra and van Neerven \cite{Zijlstra:1994x1} have pointed out that
the AB scheme described above is just one of a family of schemes
keeping $\Delta{q_{NS}}$ scale independent.
\begin{equation}
  \pmatrix{\Delta\Sigma\cr \Delta g}_a =                       
  \pmatrix{\Delta\Sigma\cr \Delta g}_{\MSbar} 
  + \frac{\alpha_s}{2\pi} \pmatrix{0 &z(x;a)_{qg}\cr
                                   0 &0             } \otimes
  \pmatrix{\Delta\Sigma\cr \Delta g}_{\MSbar},
\end{equation}
where $z_{qg}(x;a)=N_f[(2x-1)(a-1)+2(1-x)]$. The AB scheme
corresponds to $a=2$; Leader \etal.\cite{Leader:1998x1} propose yet
another scheme they call the JET scheme, in which $a=1$.  In this
scheme all hard effects are absorbed into the coefficient functions
and the gluon coefficient is as it appears in $pp{\to}JJ+X$.

The transformation between the $\MSbar$ and JET schemes is then given
by the following (suppressing the $Q^2$ dependence):
\begin{eqnarray}
  \Delta \Sigma^{(n)}_{\rm JET}
  &=&
  \Delta \Sigma^{(n)}_{\MSbar} +
  \frac{n_f\alpha_s}{2\pi} \frac{2}{n(n+1)} \Delta g^{(n)}_{\MSbar},
\\
  \Delta g^{(n)}_{\rm JET}
  &=&
  \Delta g^{(n)}_{\MSbar}.
\end{eqnarray}
For example, such a transformation indicates that the polarised
strange sea, $\Delta{s}$, will be different in the two schemes. Of
course, AB and JET are the same for $n=1$. The analogous
transformation of the coefficient functions and anomalous dimensions
from the $\MSbar$ to the AB scheme is given by replacing the factor
$2/n(n+1)$ with $1/n$.  Thus, the ABJ anomaly, far from being an
obstacle, may provide a route to parton definitions of a physically
intuitive and meaningful form.

%------------------------------------------------------------------------------%
\subsection{Small-\protect\autobf{$x$} Extrapolation}

The main problem with regard to parametrisation is the extrapolation
$x{\to}0$.  As shown by De R\'ujula \cite{DeRujula:1974x1} and later
studied by Ball and Forte,\cite{BFR} PQCD evolution leads to the
following \emph{un}-polarised small-$x$ asymptotic behaviour:
\begin{eqnarray}
  g &\sim& x^{-1}
  \sigma^{-1/2}e^{2\gamma\sigma-\delta\zeta}
  \left(1 +
  \sum_{i=1}^n \epsilon^i\rho^{i+1}\alpha_s^i \right)
\\
  \Sigma &\sim& x^{-1}
  \rho^{-1}\sigma^{-1/2}e^{2\gamma\sigma-\delta\zeta}
  \left(1 +
  \sum_{i=1}^n \epsilon_f^i\rho^{i+1}\alpha_s^i \right),
\end{eqnarray}
$\xi=\log{x_0/x}$,
$\zeta=\log{\left(\alpha_s(Q_0^2)/\alpha_s(Q^2)\right)}$,
$\sigma=\sqrt{\xi\zeta}$, $\rho=\sqrt{\xi/\zeta}$, and the
$\epsilon^i$ terms indicate $i$-th order corrections.  In the
\emph{un}-polarised case, the leading singularity is carried by
gluons, which drive the singlet quark evolution. However, all
polarised singlet anomalous dimensions are singular and therefore
gluons and quarks ``mix''. Moreover, the asymptotic predictions hold
only for \emph{non}-singular input densities: a singular starting
point is preserved. It follows that the structure functions $xF_1$ and
$F_2$ rise at small $x$ more and more steeply as $Q^2$ increases,
though, for all finite $n$, never as steeply as a power of $x$.

All other parton densities $f$ ($f=q_{NS}$, $\Delta q_{NS}$,
$\Delta\Sigma$, $\Delta g$) behave as
\begin{equation}
  f \sim
  \sigma^{-1/2}e^{2\gamma_f\sigma-\delta_f\zeta}
  \left( 1
       + \mbox{$\sum_{i=1}^n$} \epsilon_f^i\rho^{2i+1}\alpha_s^i
  \right).
\end{equation}
These last are less singular than the unpolarized singlet densities by
a power of $x$, while the higher-order corrections are more important
at small $x$ since the exponent $i+1$ is replaced by $2i+1$; because
the leading small $N$ contributions to the anomalous dimensions at
order $\alpha_s^{i+1}$ are $\left(\alpha_s/(N-1)\right)^i$ in the
unpolarized singlet case, but $N\left(\alpha_s/N^2\right)^i$ for the
non-singlet and polarized densities.  Altarelli \etal.\ obtain better
fits using a logarithmic form (rather than a power). Although this is
reminiscent of evolution effects and is compatible with Regge theory
too, no conclusions can be drawn from such results.  As a final
comment, fits generally give good overall agreement with PQCD
evolution.

%------------------------------------------------------------------------------%
\subsection{Input Sea Symmetry Assumptions}

To fit data, assumptions for the sea polarisation are usually
necessary; a common choice is flavour symmetry:
$\Delta\bar{s}=\Delta\bar{u}=\Delta\bar{d}$.  To test such a
hypothesis, Leader \etal.\cite{Leader:1998x1} note that if one allows
$\Delta\bar{s}=\Delta\bar{u}=\lambda\Delta\bar{d}$, then the data (via
$\beta$-decay couplings) fix $\Delta{q_{3,8}}$, $\Delta\Sigma$ and
$\Delta{G}$.  Thus, while
\begin{equation}
  \Delta\bar{s} = \frac{1}{6} ( \Delta\Sigma - \Delta q_8 ),
\end{equation}
and therefore $\Delta\bar{s}$ clearly does not vary with $\lambda$.

On the other hand,
\begin{eqnarray}
  \Delta u_v
  &=&
  \frac12 [ \hphantom-
    \Delta q_3
  + \Delta q_8
  - 4(\lambda-1)\Delta\bar{s}],
\\
  \Delta d_v
  &=&
  \frac12 [
  - \Delta q_3
  + \Delta q_8
  - 4(\lambda-1)\Delta\bar{s}],
\end{eqnarray}
so, \emph{valence densities are sensitive to sea assumptions}.
However, the dependence on $\lambda$ can only arise via scaling
violation and hence is weak, as seen in the analysis (indeed, it is
likely an artifact).  Results for $\Delta\bar{s}$ should not change
significantly as the input value of $\lambda$ varies, thus testing the
analysis stability.

%------------------------------------------------------------------------------%
\subsection{Input Non-Singlet Shape Assumptions}

\begin{figure}[htb]
\centering
\epsfig{figure=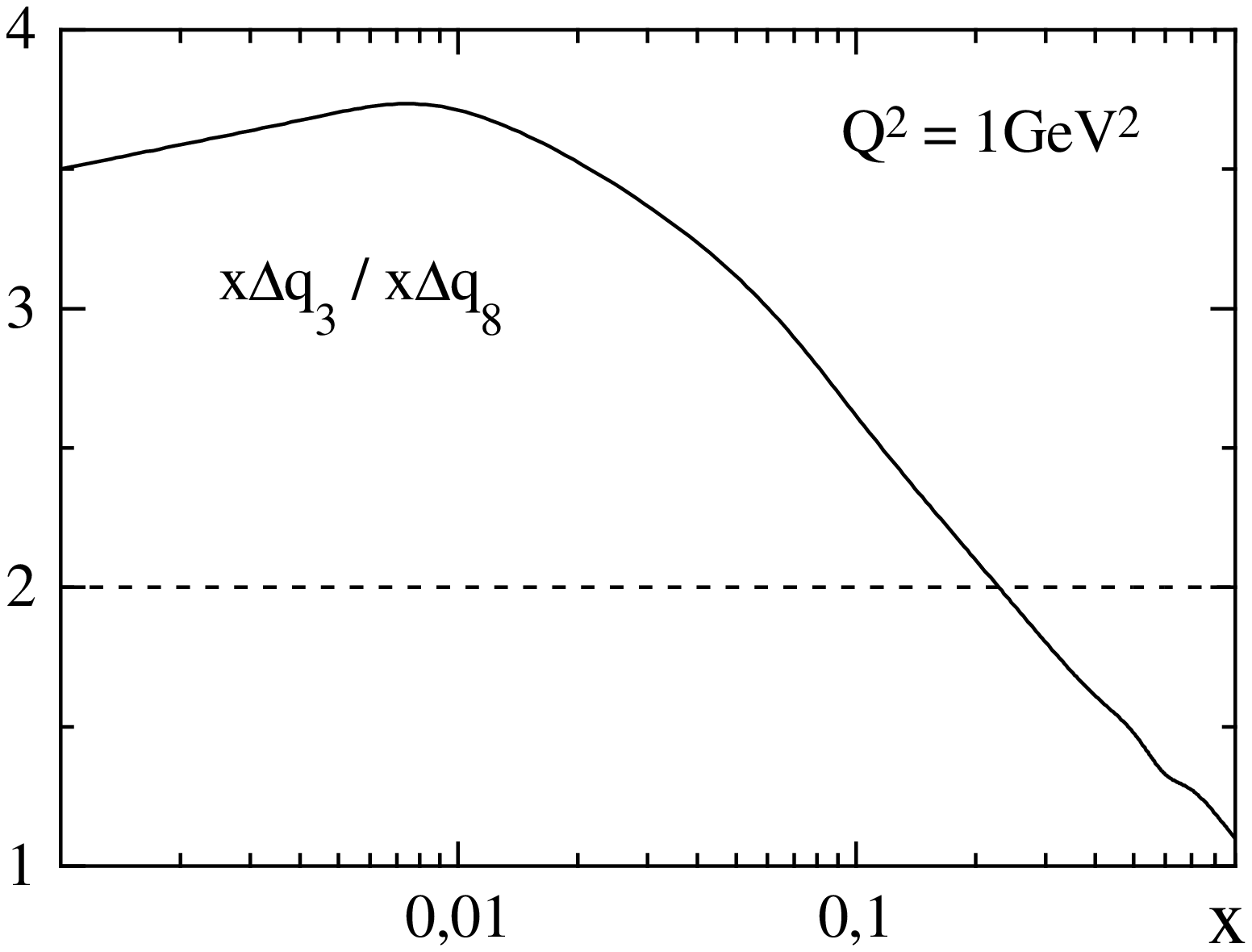,width=55mm}
\caption{\label{fig-lss4}
  The ratio $\Delta{q_3(x,Q^2)}/\Delta{q_8(x,Q^2)}$, from Leader
  \etal.\protect\cite{Leader:1998x1}}
\end{figure}

In order to reduce the number of free parameters in the fitting
procedure, a further assumption sometimes adopted is
\begin{equation}
  \Delta q_3(x,Q^2) \propto \Delta q_8(x,Q^2).
\end{equation}
While compatible with evolution (both are non-singlet densities), it
cannot be at all justified as a starting point: allowing the two
densities, $\Delta q_3(x,Q^2)$ and $\Delta q_8(x,Q^2)$, to vary
independently, significant differences are found,\cite{Leader:1998x1}
see fig.~\ref{fig-lss4} (recall the $u(x)-d(x)$ difference).  Thus,
such an assumption will certainly distort parameter values and errors
obtained.  Note too that the data do indeed constrain valence
densities well.

%------------------------------------------------------------------------------%
\subsection{Hyperon Data Input}

Together with the Bjorken sum-rule input, $a_3=F+D$, the
``hypercharge'' equivalent, $a_8=3F-D$, is also needed. The measured
baryon-octet $\beta$-decays can provide the extra information:
assuming SU$(3)_f$ symmetry, all hyperon semi-leptonic decays (HSD)
may be described moderately well in terms of the Cabibbo mixing angle
and precisely the two parameters required, $F$ and $D$.
The precision of the HSD data is better than presently needed for DIS
analyses. However, since SU(3) violation is typically of order 10\%,
one worries that the extracted values of the two parameters could
suffer the same order of shift.  

There exist SU(3)-breaking analyses
returning $F/D\simeq0.5$ (\cf\ the standard value: 0.58), but the poor
$\chi^2$ of all such fits casts doubt on their validity.  Pure SU(3)
fits to the hyperon semi-leptonic decays are also often used.  These
typically return $F/D\simeq0.575$, but again the fits are very poor:
$\chi^2\simeq2/$DoF. On the other hand, SU(3) breaking fits with only
one new parameter, give much better agreement:
$\chi^2\simeq1/$DoF.\cite{PGR-HSD} These fits return $F/D\simeq0.57$,
which is also stable with respect to the SU(3) breaking approach
adopted.

%------------------------------------------------------------------------------%
\section{Tests of Perturbation Theory}

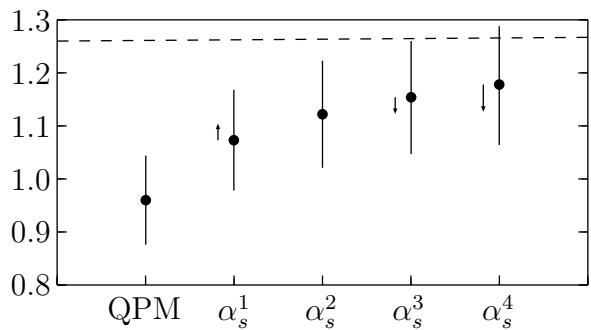
\begin{figure}[htb]
\centering
\begin{picture}(200,120)(0,-10)
\LinAxis (  0,  0)(200,  0)(6, 1, 3, 0, 0.5)
\Line    (  0,100)(200,100)
\LinAxis (  0,  0)(  0,100)(5, 1,-3, 0, 0.5)
\LinAxis (200,  0)(200,100)(5, 1, 3, 0, 0.5)
\Text    ( 33,-10)[]{QPM}
\Text    ( 67,-10)[]{$\alpha_s^1$}
\Text    (100,-10)[]{$\alpha_s^2$}
\Text    (133,-10)[]{$\alpha_s^3$}
\Text    (167,-10)[]{$\alpha_s^4$}
\Text    (-10,  0)[]{0.8}
\Text    (-10, 20)[]{0.9}
\Text    (-10, 40)[]{1.0}
\Text    (-10, 60)[]{1.1}
\Text    (-10, 80)[]{1.2}
\Text    (-10,100)[]{1.3}
\SetScale {0.2}
\SetWidth {2.5}
\SetScaledOffset(0,-800)
\DashLine(  1,1260)(999,1267){20}
\Vertex  (167, 960){10}
\Line    (167, 876)(167,1044)
\Vertex  (333,1073){10}
\Line    (333, 978)(333,1168)
\Vertex  (500,1122){10}
\Line    (500,1021)(500,1223)
\Vertex  (667,1154){10}
\Line    (667,1047)(667,1260)
\Vertex  (833,1178){10}
\Line    (833,1064)(833,1288)
\LongArrow(303,1073)(303,1103)
\LongArrow(637,1154)(637,1124)
\LongArrow(803,1178)(803,1128)
\end{picture}
\caption{\label{fig:EKplot}
  The order-by-order comparison of data and theory for the Bjorken
  sum-rule: the data points are corrected for QCD and the dashed line
  is $g_A/g_V$.  The arrows indicate the directions and relative
  magnitudes of the shifts indicated in the text.}
\end{figure}

One of the many ways to test PQCD is to compare $\alpha_s$ as
extracted in different processes.  A particularly suggestive method
used of late is to show the order-by-order agreement in, \eg, the
Bjorken sum-rule (see fig.~\ref{fig:EKplot}).  While data are
unambiguous, modulo the usual experimental uncertainties, such a plot
is misleading from a theoretical viewpoint: as commented from the
floor at this symposium, the na{\"\i}ve interpretation would
\emph{not} be convergence of the perturbation series to the correct
value, rather an imminent crossing and possible premature divergence.
Although PQCD perturbation series are generally held to be asymptotic,
this is clearly not what is being displayed here.

The problem lies in the use of a fixed-order $\alpha_s$: it is simply
incorrect to use a fixed-order extraction of $\alpha_s$ in
variable-order predictions.  In the majority of cases the perturbative
expansion displays monotonic behaviour (at least for the few known
terms), just as the Bjorken series.  Hence, as the order of
perturbation theory used for extraction increases, the value of
$\alpha_s$ obtained decreases.  Thus, taking the world mean $\alpha_s$
to be (on average) second order, a first-order extraction would
provide a relatively larger value and third and fourth orders,
progressively smaller.  Correct order-by-order comparison would then
lead to the shifts indicated in fig.~\ref{fig:EKplot} and therefore
milder convergence.

%------------------------------------------------------------------------------%
\section{Transverse Polarisation}

Transverse spin has many facets, I now turn to the status of PQCD
approaches to the long-standing puzzle of single-spin asymmetries and
also mention some recent developments in transversity.  Inequalities
are important here too, as constraints on model builders' input
densities for predictive purposes \cite{Trans-Ineq} but again, being
technical in nature, I shall not discuss them further.

%------------------------------------------------------------------------------%
\subsection{Single-Spin Asymmetries}

A most interesting aspect of transverse spin is the large amount of
SSA data: measured effects reach the level of 40--50\% in a wide range
of processes.\cite{Bravar:1998x1} Ever since Kane
\etal.,\cite{Kane:1978nd} it has been realised that at twist 2 in LO
massless PQCD such effects are zero. At NLO, the effects due to
imaginary parts of loop diagrams are found to be at most of order 1\%.

However, since the pioneering work of Efremov and
Teryaev,\cite{Efremov:1985ip} it is now well understood that twist-3
effects naturally lead to such asymmetries. To calculate the SSA in
prompt-photon production Qiu and Sterman \cite{Qiu:1991pp} have
exploited their idea of taking the necessary imaginary part from
\emph{soft propagator poles} arising in extra partonic legs inherent
to twist-3 contributions. Since then Efremov
\etal.\cite{Efremov:1995dg} have performed calculations for the pion
asymmetry, as too have Qiu and Sterman,\cite{Qiu:1998ia} and Ji
\cite{Ji:1992x2} has examined the purely gluonic contributions. The
results are all very encouraging.

%------------------------------------------------------------------------------%
\subsection{Factorisation in Higher-Twist Amplitudes}

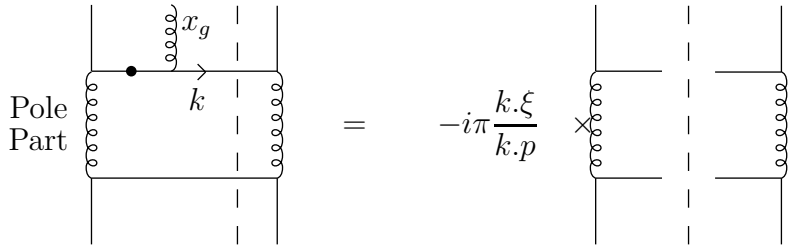
\begin{figure}[htb]
\centering
\begin{picture}(280,100)(0,120)
\multiput(0,110)(0,0){1}{
\Text  (  0, 62)[]{Pole}
\Text  (  0, 50)[]{Part}
\Line  ( 90,100)( 90, 75)
\Gluon ( 20, 75)( 20, 35){-2}{6} \Gluon( 90, 75)( 90, 35){2}{6}
\Line  ( 20, 35)( 20, 10)        \Line ( 90, 35)( 90, 10)
\Line  ( 20, 35)( 90, 35)
\DashLine( 75,100)( 75,10){7}
\Line  ( 60, 72)( 63, 75)        \Line ( 60, 78)( 63, 75)
\Text  ( 60, 65)[]{$k$}
\Line  (280,100)(280, 75)
\Gluon (210, 75)(210, 35){-2}{6} \Gluon(280, 75)(280, 35){2}{6}
\Line  (210, 35)(210, 10)        \Line (280, 35)(280, 10)
\Line  (255, 75)(280, 75)
\Line  (210, 35)(235, 35)        \Line (255, 35)(280, 35)
\DashLine(245,100)(245,10){7}
}
\Line  ( 20,210)( 20,185)        \Gluon( 50,210)( 50,185){2}{4}
\Line  ( 20,185)( 90,185)
\Vertex( 35,185){2}
\Text  (115,165)[l]{$\displaystyle=
                     \qquad-i\pi\frac{k.\xi}{k.p}\quad\times$}
\Text  ( 54,202)[]{$\quad x_g$}
\Line  (210,210)(210,185)
\Line  (210,185)(235,185)
\end{picture}
\caption{\label{fig:qfactor}
  Representation of the amplitude factorisation in the case of a soft
  external gluon line $x_g{\to}0$. The solid circle indicates the
  propagator from which the imaginary piece is extracted, and $\xi$ is
  the polarisation vector of the gluon entering the factorised
  vertex.}
\end{figure}

A difficulty in such calculations is the large number (several tens)
of PQCD diagrams encountered. Recently I have shown
\cite{Ratcliffe:1998pq} that, in the pole limit of interest, the
contributions simplify owing to a factorisation of the soft insertion
from the rest of the amplitude, see fig.~\ref{fig:qfactor}.
The remaining $2{\to}2$ helicity amplitudes are known; thus,
calculation of any such process reduces to simple products of known
helicity amplitudes with the above ``insertion'' factors (including
modified colour factors). The factorisation described, also leads to
more transparent expressions clarifying why large SSA's may be
expected: there is no reason for suppression (kinematic mismatch
\etc), beyond their higher-twist nature.

%------------------------------------------------------------------------------%
\subsection{Unravelling Higher Twist in Single Spin Asymmetries}

A complication that has emerged is that the possible mechanisms for
producing such asymmetries are numerous (even when restricted to the
purely PQCD processes described, not to mention the problem of
intrinsic $k_T$ \cite{Murgia:1998x1}). Thus, it is now important to
analyse the possible processes in which SSA's are allowed and to
identify those with differing origins, in the hope of eliminating some
candidates and finally arriving at the true source. A step in this
direction has been taken by Boros \etal.,\cite{Boros} see
table~\ref{tab:Boros}, and an interesting discussion has also been
presented at this symposium by Murgia.\cite{Murgia:1998x1}
\begin{table}[htb] \scriptsize
\centering
\vskip-4pt
\caption{\label{tab:Boros}
  Predictions for other SSA's under different hypotheses for the
  origins of the pion asymmetry, from Boros
  \etal.\protect\cite{Boros}} 
\vskip4pt
\begin{tabular}{|c|c|c|c|c|}
\hline
& \multicolumn{4}{c|}{if $A_N$ observed in $p(\uparrow)+p\to\pi+X$
  originates from \rule[-1ex]{0pt}{4ex}...}
\\
\cline{2-5}
  \parbox[c][10ex][c]{32ex}{\centering process}
& \parbox[c][10ex][c]{15ex}{\centering quark distribution function}
& \parbox[c][10ex][c]{15ex}{\centering elementary scattering process}
& \parbox[c][10ex][c]{15ex}{\centering quark fragmentation function }
& \parbox[c][10ex][c]{15ex}{\centering orbital motion and surface effect}
\\
\hline
  \parbox[c][9ex][c]{32ex}{\centering $l + p(\uparrow) \to l + 
  \left(\begin{array}{c}\pi^\pm\\ K^+ \end{array}\right) + X$}
& \parbox[c][9ex][c]{15ex}{\centering $A_N=0$    \\ wrt jet axis}
& \parbox[c][9ex][c]{15ex}{\centering $A_N=0$    \\ wrt jet axis}
& \parbox[c][9ex][c]{15ex}{\centering $A_N\neq0$ \\ wrt jet axis}
& \parbox[c][9ex][c]{15ex}{\centering $A_N=0$    \\ wrt jet axis}
\\
\cline{2-5}
  \parbox[c][9ex][c]{32ex}{\centering current fragmentation region\\ 
  for large $Q^2$ and large $x_B$}
& \parbox[c][9ex][c]{15ex}{\centering $A_N\neq0$ \\ wrt $\gamma^\star$ axis}
& \parbox[c][9ex][c]{15ex}{\centering $A_N=0$    \\ wrt $\gamma^\star$ axis}
& \parbox[c][9ex][c]{15ex}{\centering $A_N\neq0$ \\ wrt $\gamma^\star$ axis}
& \parbox[c][9ex][c]{15ex}{\centering $A_N=0$    \\ wrt $\gamma^\star$ axis}
\\
\hline
  \parbox[c][13ex][c]{32ex}{\centering $l + p(\uparrow) \to l +
  \left(\begin{array}{c} \pi^\pm\\ K^+ \end{array}\right) + X $ \\[1mm] 
  target fragmentation region\\
  for large $Q^2$ and large $x_B$}
& \parbox[c][13ex][c]{15ex}{\centering $A_N\neq0$}
& \parbox[c][13ex][c]{15ex}{\centering $A_N=0$}
& \parbox[c][13ex][c]{15ex}{\centering $A_N\neq0$}
& \parbox[c][13ex][c]{15ex}{\centering $A_N=0$}
\\
\hline
  \parbox[c][12ex][c]{32ex}{\centering $p + p(\uparrow) \to 
  \left(\begin{array}{c} l\bar{l}\\ W^\pm \end{array}\right) + X$\\[1mm]
  $p(\uparrow)$ fragmentation region}
& \parbox[c][12ex][c]{15ex}{\centering $A_N\neq0$}
& \parbox[c][12ex][c]{15ex}{\centering $A_N\approx0$}
& \parbox[c][12ex][c]{15ex}{\centering $A_N=0$}
& \parbox[c][12ex][c]{15ex}{\centering $A_N\neq0$}
\\
\hline
\end{tabular}
\end{table}

%------------------------------------------------------------------------------%
\subsection{Transversity}

\begin{figure}[htb]
\centering
\epsfig{figure=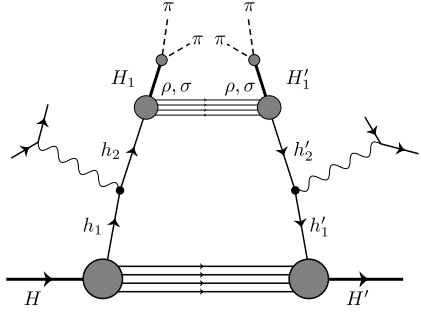,width=55mm}
\caption{\label{fig-jjt1}
  Diagram for double-pion production via $\sigma$, $\rho$ in DIS.}
\end{figure}

Despite the advantage of being twist-2 and therefore unsuppressed,
transversity cannot contribute to inclusive DIS as it requires quark
helicity flip. However, Jaffe \etal.\cite{Jaffe:1998hf} have recently
proposed its measurement via twist-two quark-interference
fragmentation functions, see fig.~\ref{fig-jjt1}.
This is remeniscent of the so-called handedness property but appears
more direct to interpret.

FSI between, \eg, $\pi^+\pi^-$, $K\overline{K}$, or $\pi{}K$, produced
in the current fragmentation region in DIS on a transversely polarised
nucleon may probe transversity. The point is that the pions may be
produced through intermediate resonant states:
$\sigma[(\pi\pi)^{I=0}_{l=0}]$ and $\rho[(\pi\pi)^{I=1}_{l=1}]$
produced in the current fragmentation region. Two new interference
fragmentation functions can then be defined: $\hat{q_I}$,
$\delta\hat{q_I}$; the subscript $I$ stands for interference.  The
final asymmetry, requiring no model input, has the following form:
\cite{Jaffe:1998hf}
\begin{eqnarray}
  {\cal A}_{\bot\top}
  \equiv
  \frac{d\sigma_\bot-d\sigma_\top}{d\sigma_\bot+d\sigma_\top}
  &=&
  -\frac{\pi}{4} \frac{\sqrt6(1-y)}{[1+(1-y)^2]}
  \cos\phi \sin\delta_0 \sin\delta_1 \sin\left(\delta_0-\delta_1\right)
\nonumber\\
  &&
  \hspace*{1ex}
  \times \; 
  \frac{\sum_a e_a^2 \delta q^a(x)\, \delta \hat{q}_I^a(z)}
  {\sum_a e^2_a q_a(x)
  \left[ \sin^2\delta_0 \hat{q}_0^a(z)
  + \sin^2\delta_1 \hat{q}_1^a(z) \right]},
\end{eqnarray}
where $\hat{q}_0$ and $\hat{q}_1$ are spin-averaged fragmentation
functions for the intermediate $\sigma$ and $\rho$ states, respectively.

Note that the target only need be polarised; the asymmetry is obtained
either by flipping the nucleon spin or via the azimuthal asymmetry.
This approach can also be extended to generate a (double) helicity
asymmetry, which could probe valence-quark spin
densities.\cite{Jaffe:1998pv}

%------------------------------------------------------------------------------%
\section{Orbital Angular Momentum}

It has long been well understood that angular-momentum conservation in
parton splitting processes implies non-trivial $Q^2$ evolution of
partonic orbital angular momentum (OAM),\cite{Ratcliffe:1987dp} in
line with that of the gluon spin. Ji \etal.\cite{Ji:1996x1} have shown
that this leads to asymptotic sharing of total angular momentum
identical to that of the linear momentum fraction. Recently, Teryaev
\cite{Teryaev:1998x2} has rederived the PQCD evolution equations for
OAM in a semi-classical approach in terms of the spin-averaged and
spin-dependent kernels.

Generation of OAM, balancing the gluon spin, accompanies $q{\to}qg$
splitting. The net effect is obtained subtracting the probabilities of
a gluon with negative and positive helicities. The same combination
(modulo sign) appears in the $gq$ spin-dependent kernel, with momentum
fraction $1{-}x$:
\begin{equation}
  P^{LS}_{qq}(x) + P^{LS}_{gq}(1-x) = -\Delta P_{gq}(1-x).
\end{equation}
The trick is that the ratio of the quark and gluon OAM can be found
via classical reasoning. Before splitting, suppose the quark momentum
has only a $z$ component, the final parton momenta are in the $x$-$z$
plane.  By momentum conservation, the $x$ components of $q$ and $g$
momenta are equal (up to a sign) and the $z$ components of OAM are
thus
\begin{equation}
  L_z^q= P_x r^q_y, \qquad L_z^g= -P_x r^g_y,
\end{equation}
where the spatial non-locality of quark and gluon production,
$r^{q,g}$, has been introduced. OAM $x$ components are also generated:
\begin{equation}
  L_x^q= -P^q_z r^q_y, \qquad L_x^g= -P^g_z r^g_y.
\end{equation}
Conservation of the $x$ component of OAM, $L_x^q=-L_x^g$, leads to
\begin{equation}
  \frac{r^q_y}{r^g_y} = -\frac{P_z^g}{P_z^q},
\end{equation}
and substitution into the $L_z$ eq.\ finally gives the partition:
\begin{equation}
  \frac{L_z^q}{L_z^g} = \frac{P_z^g}{P_z^q} = \frac{1-x}{x},
\end{equation}
precisely as Ji \etal.\cite{Ji:1996x1} found by explicit calculation.
Notice also that the whole problem of defining the relevant operators
has been neatly circumvented.

%-------------------------------- Conclusions ---------------------------------%
\section{Conclusions}

In conclusion then, all theoretical aspects of spin physics continue
to benefit from interest and study.  Moreover, the rewards for the
effort put into this sector are an ever-deeper understanding of
hadronic structure (and may indeed represent one of the few real keys)
while the various phenomenological puzzles are steadily coming under
control.

Areas where there is more to be learnt are transverse spin (including
transversity) and orbital angular momentum.  The former, being linked
to hadronic mass scales may provide important clues to the nature of
chiral-symmetry breaking while there is also still much to explain of
the known phenomenology, and transversity may yet have a r\^ole in
single-spin asymmetries.  OAM is as intriguing as it is elusive
experimentally; its contribution the proton spin is yet to be measured
and, were it to be found large, one would like to understand the
implications for the standard SU(6) picture of the nucleon.

%-------------------------------- Bibliography --------------------------------%

%---------------------------------- The End -----------------------------------%
\end{document}